\documentclass[12pt,epsf]{article}

\usepackage{scicite}

\input psfig.sty

\def\gsim{\mathrel{\rlap{\lower 4pt \hbox{\hskip 1pt $\sim$}}\raise 1pt
\hbox {$>$}}}
\def\lsim{\mathrel{\rlap{\lower 4pt \hbox{\hskip 1pt $\sim$}}\raise 1pt
\hbox {$<$}}}

\newenvironment{sciabstract}{%
\begin{quote} \bf}
{\end{quote}}

\newcounter{lastnote}

\title{Protostar Formation in the Early Universe} 
\author{
Naoki Yoshida,$^{1}$ 
Kazuyuki Omukai,$^{2}$ 
Lars Hernquist$^{3}$ \\
\\
\normalsize{$^{1}$Department of Physics, Nagoya University, 
Furocho, Chikusa, Nagoya, 
Aichi 464-8602, Japan}\\ 
\normalsize{$^{2}$National Astronomical Observatory of Japan, 
Osawa, Mitaka, Tokyo 181-8588, Japan}\\
\normalsize{$^{3}$Harvard-Smithsonian Center for Astrophysics,}\\
\normalsize{60 Garden Street, Cambridge, MA02138}\\
\\
\normalsize{$^\ast$To whom correspondence should be addressed; E-mail:
nyoshida@a.phys.nagoya-u.ac.jp.} \\
\normalsize{{\bf Science, 321, 669 (2008)}}
}

\date{}

\begin{document} 


\baselineskip24pt


\maketitle


\begin{sciabstract}

The nature of the first generation of stars in the Universe remains largely unknown.
Observations imply the existence of massive primordial stars
early in the history of the universe, and the standard theory for 
the growth of cosmic structure predicts that structures grow hierarchically 
through gravitational instability. 
We have developed an {\it ab initio} computer simulation of the
formation of primordial stars that
follows the relevant atomic and molecular processes in a primordial gas
in an expanding universe. 
The results show that primeval density fluctuations left over from the Big Bang
can drive the formation of a tiny protostar with a mass of just 
one percent that of the Sun. 
The protostar is a seed for the subsequent formation of a massive primordial star.

\end{sciabstract}

Large ground-based telescopes have discovered distant astronomical
objects such as galaxies and quasars\cite{SDSS, Iye} that were in
place when the Universe was less than 1 billion years old,
or about $5\%$ of its current age.  Moreover, these
studies have shown that other luminous objects must have been present
even earlier.  For example, the most distant known quasar, SDSS-J1148,
contains substantial amounts of heavy elements such as carbon, oxygen,
and iron as well as dust grains\cite{Walter}.  These heavy elements
are not of cosmic origin, but must have been formed earlier in massive
stars before being expelled by supernovae and stellar winds, and then
incorporated into the material that later condensed to produce this
quasar.

Recently, stars with extremely low heavy element content 
were discovered in the halo of our Galaxy\cite{C02, F04}. 
The observed elemental abundance
patterns indicate several possibilities
for the nature of their {\it ancestors}\cite{Iwamoto, Komiya}.
One interesting scenario is
that supernova explosions of massive primordial stars enriched the
parent gas clouds, from which these halo stars were born.

Theoretical analyses hold promise for revealing the process
of primordial star formation for two main reasons: (i) the initial
conditions, as determined cosmologically, are well-established, so
that statistically equivalent realizations of a standard model
universe can be accurately generated, and (ii) the important basic
physics such as gravitation, hydrodynamics, and atomic and molecular
processes in a hydrogen-helium gas are understood. 
Other complications that plague
investigations of star formation in the local Universe, 
such as the presence of strong 
magnetic fields or heavy elements, 
can be neglected at these early times.

Here, we report super-computer simulations of the development of cosmic 
structure in the early Universe and the formation of primordial stars.  Our
simulations achieve a dynamic range in spatial scale of $\sim
10^{13}$, resolving small-scale structures having sizes of a fraction
of a solar radius ($\sim 10^{10} {\rm cm}$) within cosmological
volumes hundreds of kiloparsecs in length ($\sim 10^{23} {\rm cm}$).
The smallest length scale, the so-called local Jeans length set by
the action of gravity and hydrodynamic pressure, is fully resolved
throughout the simulation volume at all times.

We do not assume any {\it a
priori} equation of state for the gas. The thermal and chemical
evolution of the gas is determined fully by molecular and atomic
processes, which are treated in a direct, self-consistent manner.
The spatial resolution and the accurate implementations of the 
physical processes allow us to follow the collapse of a gas to 
stellar densities, and thus our calculations 
offer a detailed picture of how the first cosmological objects
-- primordial protostars -- form from chemically pristine gas.

We set up cosmological initial conditions such
that the statistical properties of the density and velocity fields are
matched to those given by the standard model of the Universe\cite{wmap},
according to which the energy density is dominated by dark energy and
cold dark matter.  We follow the gravitational collapse of dark matter
and the hydrodynamics of primordial gas through simulations of
cosmic structure formation.  Below, we provide details on one 
simulation, which followed the evolution of dark matter and gas
in a cube 200 comoving kiloparsecs on a side.  We focus our attention
on a gravitationally bound dark matter halo which formed in this volume
at an epoch when the cosmological redshift was $z = 14$ (Fig.~1).  
The mass of this halo, a half million solar masses, and the physical conditions within
it are particularly conducive for it to host a primordial star\cite{BCL99, ABN, BCL02}.  
The gas within this halo had a temperature of $\sim 1000$
Kelvin and a small mass of hydrogen molecules ($\sim 10^{-4}$ in
number fraction) had already formed, enabling efficient radiative
cooling.

Through the action of radiative cooling, a star-forming gas cloud
collected in the host dark halo. We tracked the subsequent thermal 
and chemical evolution of the primordial gas cloud for
more than 20 decades in density up to the epoch of protostar formation.  
We accounted for:
(i) the chemistry at both
low and high densities, including molecular hydrogen formation, 
(ii) transfer of molecular line photons and accompanying radiative cooling,
(iii) transfer of collisionally-induced continuum radiation and
resulting radiative cooling.  In the final phase of the collapse, the
temperature increased adiabatically as a result of the absence of radiative
and chemical cooling.  
This continued until the contraction of the central
part was halted by strong thermal pressure. At this time, strong
hydrodynamic shocks formed, marking the moment of the formation of
a protostar.
We could not follow the evolution after this epoch 
accurately without implementing radiative effects from the
post-shock high temperature gases, and thus we stopped 
the simulation at this point.

The structure in and around the newly formed protostar
wass rather complex (Fig.~1 D).
At this time, there were substantial variations in density and temperature 
even in the innermost 10 solar-radii region.
Clearly, the primordial protostar was not simply a sphere
surrounded by a single accretion shock. 

The cloud evolution was dictated by several important
physical processes:
First, a fully molecular cloud with mass $\sim 1
M_{\odot}$ formed\cite{ABN, ON98, Y06} when the gas density was
sufficiently high ($> 10^{8} {\rm cm}^{-3}$) that three-body chemical
reactions converted nearly all the hydrogen into molecules.  
Efficient cooling by rovibrational transitions of hydrogen molecules 
caused the first small dip at a radius of $R\sim 10^{16}\; {\rm cm}$ 
in the radial temperature profile (Fig.~2). 
When the dense, molecular part contracted further, it became optically thick to
hydrogen molecular rovibrational lines, and then the efficiency of 
radiative cooling saturated (dot-dashed lines in Fig.~2).  
At still higher densities, frequent
collisions between hydrogen molecules led to efficient emission of
continuum radiation via collision-induced emission.  By this
rapid cooling occurring at densities greater than $\sim 10^{14}\;{\rm cm}^{-3}$, a
small central part with $\sim 0.1 M_{\odot}$ 
cooled efficiently to form a flattened disk-like structure. 
In the flattened gas cloud, radiation escaped 
preferentially in the direction where contraction was fastest,
because the velocity gradient was large
and also the continuum optical depth was small in this direction 
[supporting online material (SOM) text and Fig. S1, S2]. 
This combined effect of gravitational contraction and 
direction-dependent radiative cooling accelerated the deformation
of the cloud core to a disk structure. 
The disk structure had a radius of $\sim 10^{13} {\rm cm}$ and a mass of $\sim 0.1
M_{\odot}$, where the cooling time and the local dynamical 
time are comparable. While the innermost portion further
contracted slowly, spiral density waves were excited, yielding two arms 
(see the bottom-right panel of Fig.~1).  

When the central density reached $n\sim 10^{18}\;{\rm cm}^{-3}$, 
the gas became completely optically thick to continuum radiation, and at
this point radiative cooling no longer operated efficiently
(long dashed lines in Fig.~2).
Further collapse and the associated dynamical heating triggered 
full-scale dissociation of hydrogen molecules in the central part
(see also Fig.~3).  
After most of the hydrogen molecules were collisionally dissociated,
the gas could not lose its thermal energy either radiatively or 
by dissociating molecules. 
The resulting effective equation of state became progressively more
stiff, making the gas cloud resist gravitational deformation and
fragmentation\cite{Matsumoto00, Omukai05}.  
The gas then contracted adiabatically, 
and its temperature quickly increased above several thousand Kelvin,
while the density exceeded $n\sim 10^{20}\;{\rm cm}^{-3}$.
The strong thermal pressure finally stopped the gravitational collapse
and hydrodynamic shocks were generated (solid lines in Fig.~2).
We define a constant density, atomic gas core as a protostar that
is pressure-supported. At the final output time, a protostar 
formed with a mass of just 0.01 solar masses. 
It had an initial radius of $\sim 5\times 10^{11}$ cm,
similar to that of present-day protostars in theoretical
calculations\cite{Larson69}.  The central particle number density of
the protostar was $\sim 10^{21} {\rm cm}^{-3}$ and the temperature was
well above 10,000 K.

At the time of protostar formation, the central temperature was 
so high that almost all the molecules were collisionally
dissociated within an enclosed mass of 0.01 $M_{\odot}$ (Fig.~3). 
A slight degree of ionization was also seen in the
innermost high-pressure part of the atomic core.
There was a small variation in the atomic hydrogen fraction at 
$\sim 0.1 M_{\odot}$;
a small fraction of hydrogen molecules were dissociated
when the gas temperature reached $\sim 2000$K\cite{ON98, Ripamonti02}, 
but the molecular fraction increased again
when the density increased and efficient continuum cooling 
brought the gas temperature temporarily below 2000 K
(see the second temperature dip at $R \sim 10^{13} {\rm cm}$ in Fig.~2).

The protostar accretes the ambient gas in a complicated way (Fig.~3 bottom).
In the direction vertical to the disk plane, gas falling in
at a velocity exceeding 10 km/sec was suddenly stopped at the
location of a shock at $M_{\rm enclosed} \sim 0.01 M_{\odot}$. 
Fig.~3 also shows the degree of rotational support of the gas,
defined as $f_{\rm rot} = (L/r)/v_{\rm Kep}$, where $L$ is the specific
angular momentum of the gas within radius $r$, and $v_{\rm Kep}$
is the Keplerian velocity at that radius. 

Within a mass of $\sim 0.1 M_{\odot}$, two spiral arms rotated rapidly,
and the outer part ($\sim 0.05-0.1 M_{\odot}$)
appeared nearly centrifugally supported, 
whereas the central part had gravitationally collapsed.
The central core lost part of its angular momentum
via gravitational torques exerted by non-axisymmetric
perturbations. The newly-born protostar was 
supported by both thermal pressure and rotation.
The overall velocity structure is characteristic of a collapsing
gas with an initially slow rotation,
as reported in previous studies of both
present-day and primordial star-formation\cite{Matsumoto03, Machida08}.

A long-standing question is whether a primordial gas cloud 
such as that studied here experiences vigorous fragmentation 
by thermal instability during its evolution.  
In our simulation, a single small proto-stellar
core formed first and the central part did not fragment
into multiple objects before protostar formation.  
At all phases, the locally estimated growth time for isobaric
perturbations was longer than, or only comparable to, the local dynamical time for
collapse. Hence, the cloud core did not fragment 
by thermal instability, but instead its collapse accelerated.

It has been suggested that the central part of primordial gas clouds may 
break up into smaller clumps later during
the subsequent accretion phase\cite{Clark08, Machida08}.
We have examined a core fragmentation model of \cite{Matsumoto03} by measuring 
$\Omega\;t_{\rm dyn}$ where $\Omega$ is the mean angular velocity 
and $t_{\rm dyn} = 1/\sqrt{4\pi G \rho}$ 
is the local dynamical time.
The central $\sim 0.01 M_{\odot}$ portion had a value
of $\Omega\; t_{\rm dyn} = 0.25$, which is large and 
close to the critical value for fragmentation.
Thus the formation of multiple stellar
systems may be possible, although not very likely, 
during later accretion phases.

The instantaneous gas mass accretion rate 
at the time of protostar formation was as large as
0.01-0.1 solar masses per year within the innermost 10 $M_{\odot}$.  
If the gas in the inner portion accreted
efficiently, the protostar would quickly grow to be as massive as 10
solar masses within a thousand years\cite{BL04}. Even if multiple stellar seeds
formed, there would be at least one main accreting protostar.  
A detailed proto-stellar calculation for a similarly large accretion rate 
predicts that the mass of the star when it lands on the main sequence 
will be $\sim 100 M_{\odot}$\cite{OP03, Y06}.

Feedback effects, in particular those from
ionizing photons emitted by the protostar, work to evaporate the
surrounding gas and to halt gas accretion. A semi-analytic calculation
including this radiative feedback and the effect of rotation shows that the final 
stellar mass can still be greater than a few tens of solar masses in a 
reasonable parameter space of the model\cite{MT08}.  
If instead mass accretion is unimpeded throughout the
star's evolution, the final stellar mass can be very large, 
possibly exceeding a few hundred solar masses\cite{OP03, BL04}.
Such very massive stars ionize a large volume of the surrounding gas. 
Because of the different thermal evolution of an initially ionized gas\cite{UI00},
second generation primordial stars formed
under such conditions are predicted to be several tens of solar masses
\cite{Johnson06, YOH}.
Therefore, in either case, our model provides
a viable scenario for the early chemical enrichment in the Universe
by massive primordial stars\cite{Iwamoto}, 
which is necessary for the formation of later populations of ordinary 
stars. 

The basic properties of the particular star-forming cloud we simulate, 
such as physical size and mass, 
are characteristic for
cosmological primordial gas clouds,
and the object is indeed similar in many aspects to those found in previous 
works\cite{ABN, BCL02, Oshea07}.
The final evolution of the central high-density part will likely 
be affected by its angular momentum content\cite{Machida08}. 
However, because the bulk of the cloud core
is assembled from material with
low angular momentum\cite{Y06}, 
it generally has a slow initial spin, and thus the evolution of prestellar
gas is expected to be similar to what is presented here.
Our simulation thus offers a complete
picture of how a primordial protostar may have formed from tiny cosmological
density fluctuations. 
Primordial star formation for different cosmological 
models has been explored\cite{Gao07, Spolyar08}.
The particle properties of dark matter may be another 
important factor in star-formation in the early universe.

\clearpage
\begin{figure}
\begin{center}
\psfig{file=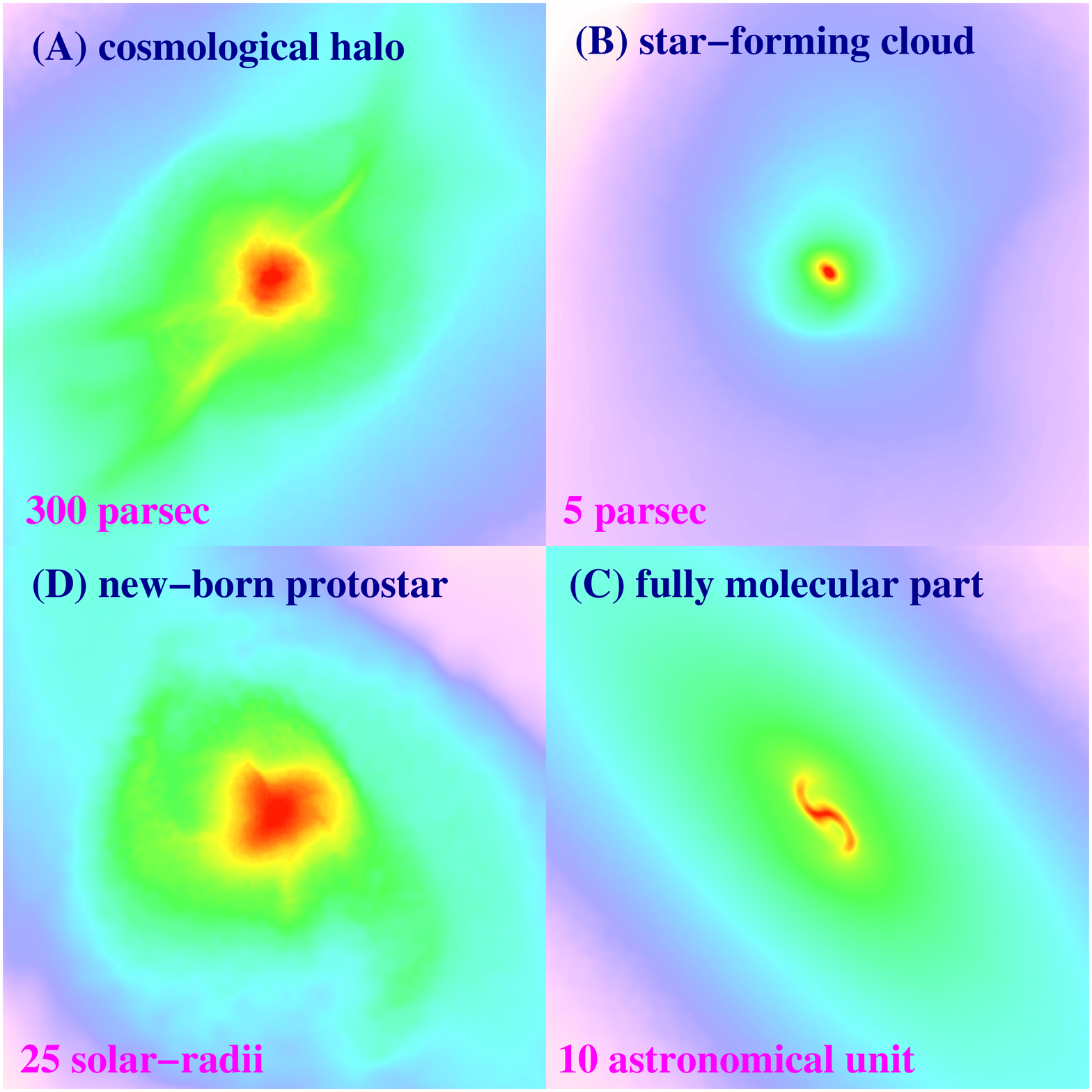,width=1\textwidth}
\caption{ Projected gas distribution around
the protostar. Shown regions are, from top-left, clockwise, 
(A) the large-scale gas distribution around the cosmological halo
(300 pc on a side), (B) a self-gravitating, star-forming cloud (5 pc on a side), 
(C) the central part of the fully molecular core (10 astronomical units on a side),
and (D) the final protostar (25 solar-radii on a side).
We use the density-weighted temperature to color (D),
to show the complex structure of the protostar.}
\end{center}
\end{figure}

\clearpage
\begin{figure}
\begin{center}
\psfig{file=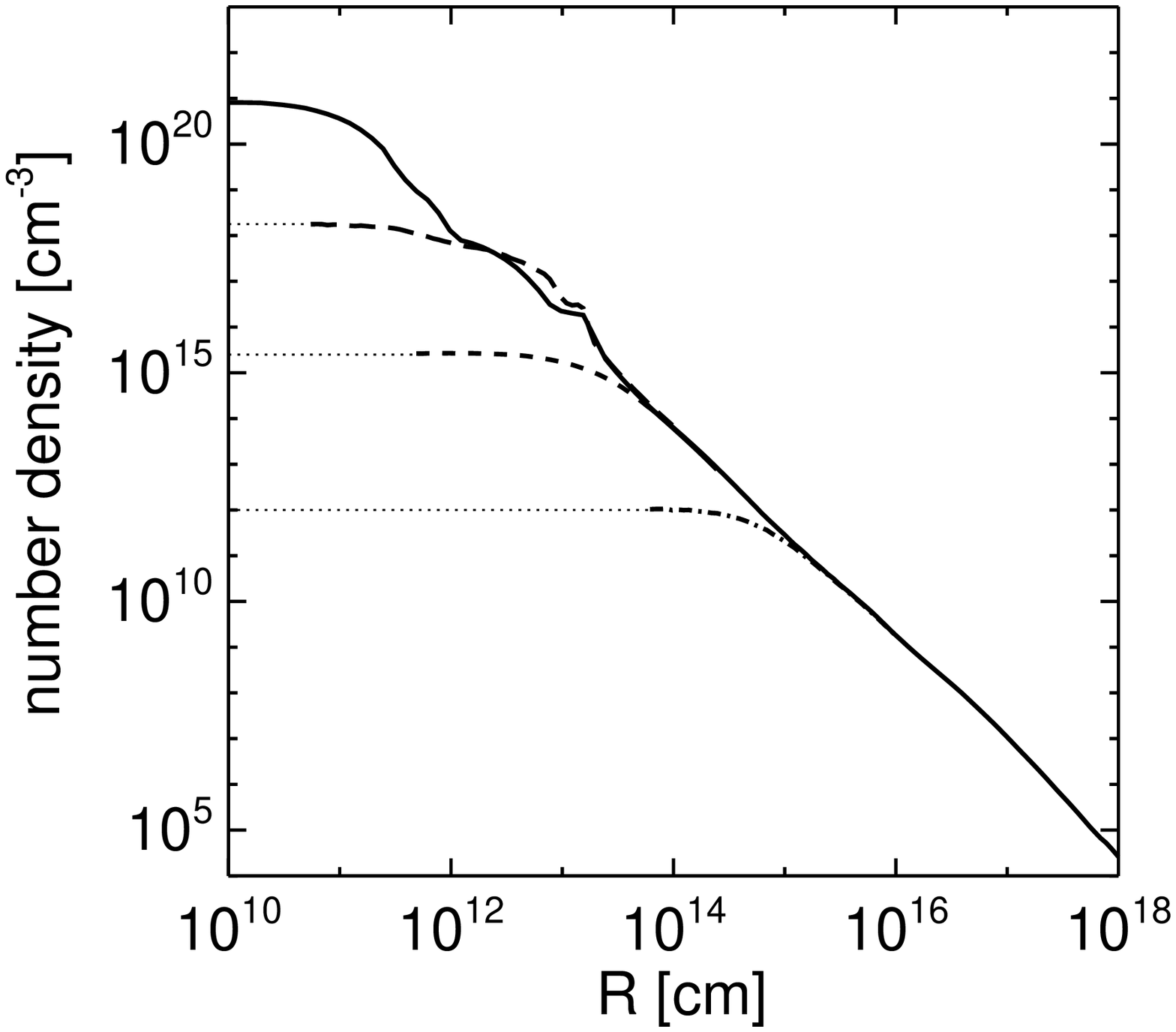,width=0.63\textwidth}
\psfig{file=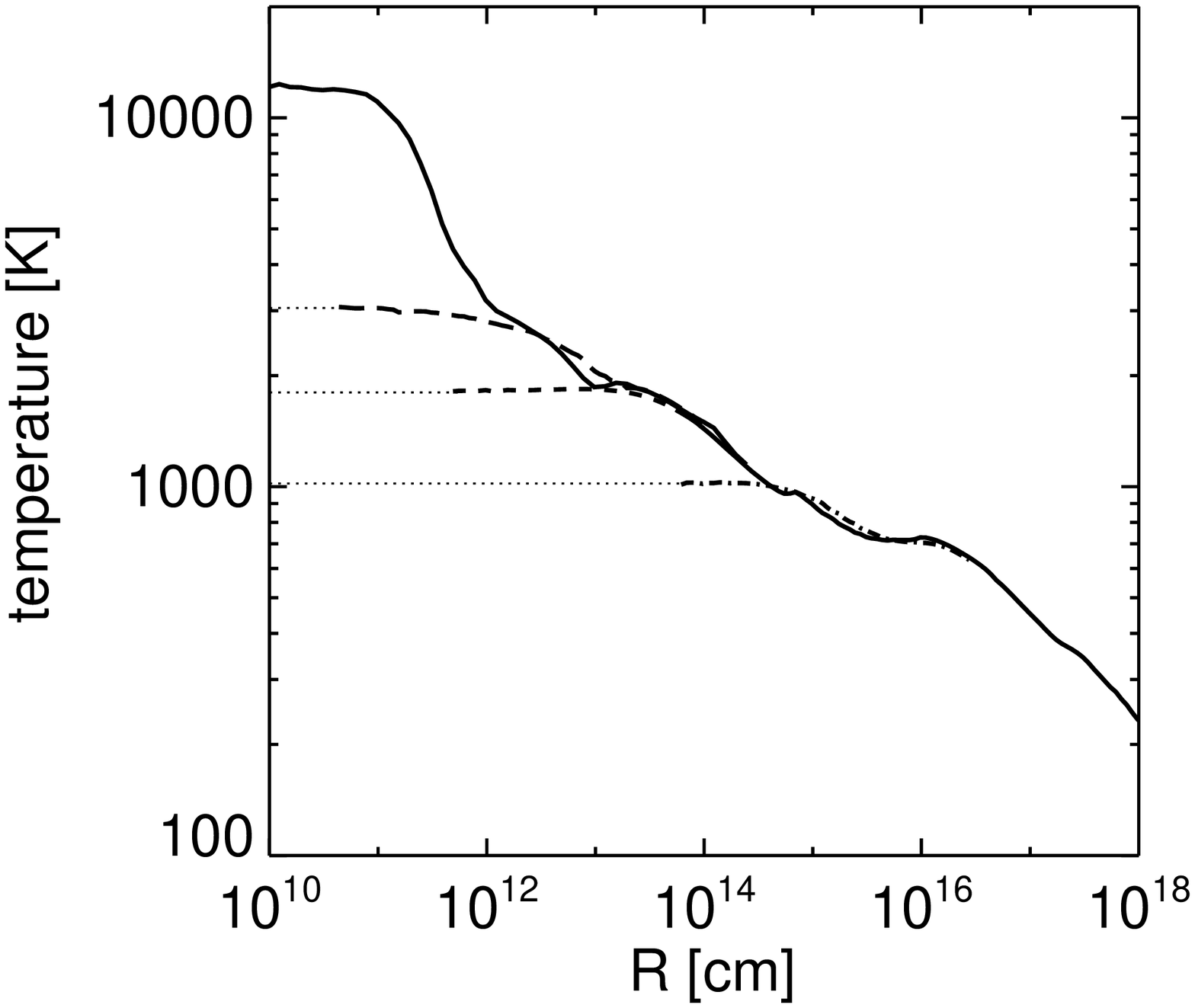,width=0.63\textwidth}
\caption{Evolution of spherically-averaged radial density 
profile (top) and temperature profile (bottom) around the protostar.
Epochs are shown when the central core
became optically thick to molecular lines (dot-dashed lines),
when cooling by collision-induced emission kicked-in (short dashed lines),
when the core became optically thick to continuum (long dashed lines),
and when full-scale dissociation was completed and 
a pressure-supported core formed (solid lines).
\label{fig:profile}}
\end{center}
\end{figure}

\clearpage
\begin{figure}
\begin{center}
\psfig{file=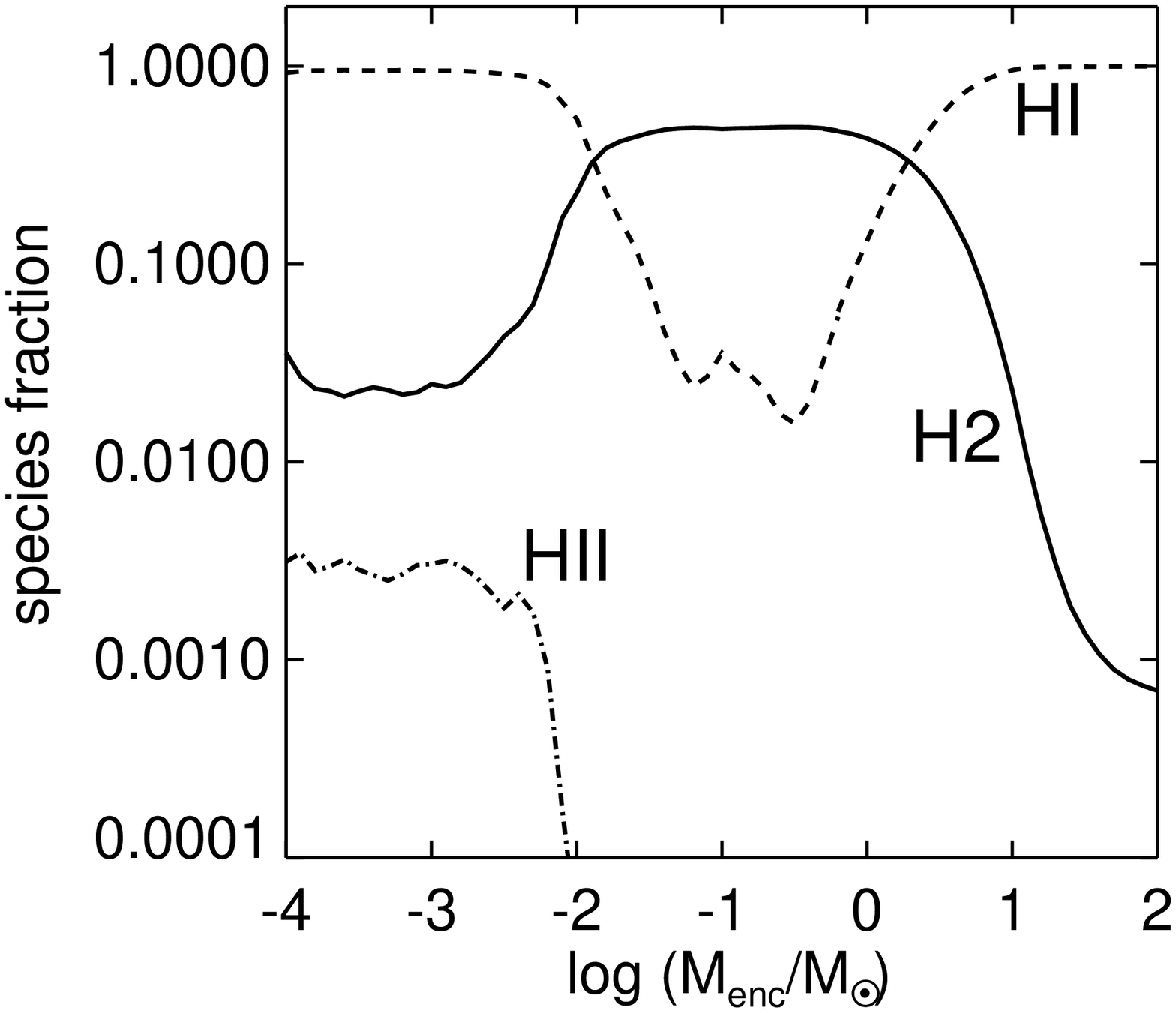,width=0.63\textwidth}
\psfig{file=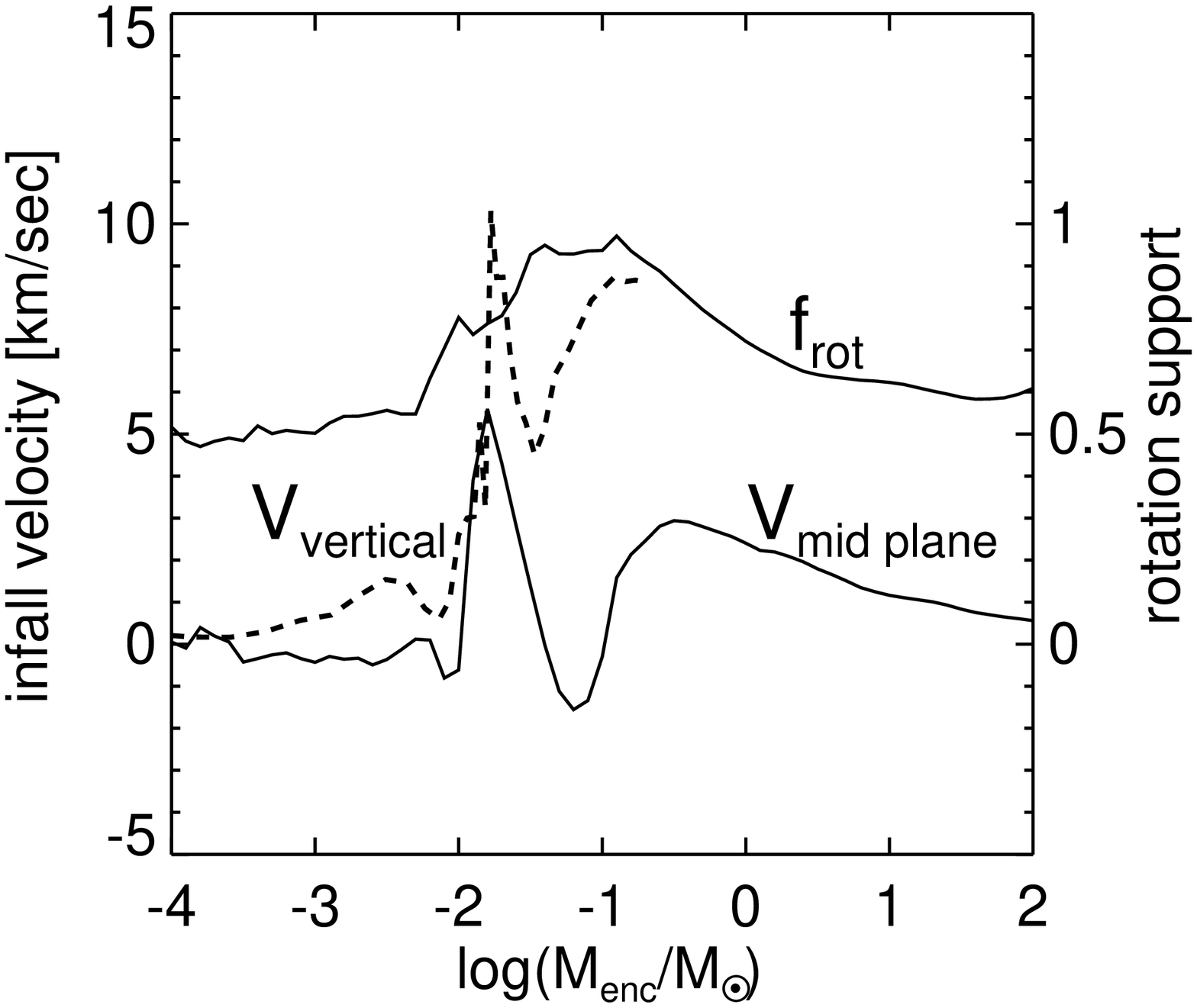,width=0.63\textwidth}
\caption{
Structure of the protostar.
(Top panel) The number fractions of atomic hydrogen (dashed line), 
molecular hydrogen (solid line), and ionized hydrogen (dot-dashed line).
(Bottom panel) The gas infall velocity
in one direction perpendicular to the disk plane (dashed line)
and an azimuthally averaged inflow velocity at the midplane (solid line).
The thin solid line shows the degree of rotational support
as defined in the text.
\label{fig:structure}}
\end{center}
\end{figure}
\end{document}